# Fuzzy Approach to Extract Pertinent Information from Web Resources

Radhouane Boughamoura[1] , Mohamed Nazih Omri[2] , Habib Youssef [3]

[1] Département d'Informatique, FSM,
Route de Kairouan, 5000 Monastir, Tunisia
bradhouane2 @yahoo.fr

[2] Unité PRINCE, Département de Technologie, IPEIM
Rue Ibn El Jazzar, 5000 Monastir, Tunisia
nazih.omri@ipeim.rnu.tn

[3]Unité PRINCE, Département d'Informatique, ISITC
Hammam Sousse, Tunisia
habib.youssef@fsm.rnu.tn

**Abstract.** Recent work in machine learning for *information extraction* has focused on two distinct sub-problems: the conventional problem of filling template slots from natural language text, and the problem of *wrapper induction*, learning simple extraction procedures ("wrappers") for highly structured text such as Web pages. For suitable regular domains, existing wrapper induction algorithms can efficiently learn wrappers that are simple and highly accurate, but the regularity bias of these algorithms makes them unsuitable for most conventional information extraction tasks. This paper describes a new approach for wrapping semi-structured Web pages. The wrapper is capable to learn how to extract relevant information from web resources on the basis of user supplied examples. It is based on inductive learning techniques as well as fuzzy logic rules. Experimental results show that our approach achieves noticeably better precision and recall coefficient performance measures than SoftMealy, which is one of the most recently reported wrappers capable of wrapping semi-structured web pages with missing attributes, multiple attributes, variant attribute permutations, exceptions, and typos.

## 1  Introduction

*Information extraction* (IE) is the problem of converting text such as newswire articles or Web pages into structured data objects suitable for automatic processing. An example domain, first investigated in the Message Understanding Conference (MUC) [3], is a collection of newspaper articles describing terrorist incidents in Latin America. Given an article, the goal might be to extract the name of the perpetrator and victim, and the instrument and location of the attack.

Research with this and similar domains demonstrated the applicability of machine learning to IE [25, 16, 27].

The increasing importance of the Internet has brought attention to all kinds of automatic document processing, including IE. And it has given rise to problem domains in which the kind of linguistically intensive approaches explored in MUC are difficult or unnecessary. Many documents from this realm, including email, Usenet posts, and Web pages, rely on extra-linguistic structures, such as HTML tags, document formatting, and ungrammatical stereotypic language, to convey essential information. Much recent work in IE, therefore, has focused on learning approaches that do not require linguistic information, but that can exploit other kinds of regularities. To this end, several distinct rule-learning algorithms [26, 19, and 9] and multi-strategy approaches [7] have been shown to be effective. Recently, statistical approaches using hidden Markov models have achieved high performance levels [28, 10, and 8].

At the same time, work on information integration [1, 11] has led to a need for specialized wrapper procedures for extracting structured information from database-like Web pages. Recent research [21, 20, 2, and 12] has shown that wrappers can be automatically learned for many kinds of highly regular documents, such as Web pages generated by CGI scripts. These *wrapper induction* techniques learn simple but highly accurate contextual patterns, such as "to retrieve a URL, extract the text between "<A href=" and ">". Wrapper induction is harder for pages with complicated content or less rigidly structured formatting, but recent algorithms [2, 12] can discover small sets of such patterns that are highly effective at handling such complications in many domains.

In this paper, we demonstrate that our fuzzy approach can be used to perform extraction in traditional (natural text) domains. We describe fuzzy approach, a trainable IE system that performs information extraction in both traditional (natural text) and wrapper (machine-generated or rigidly structured text) domains. Fuzzy approach learns extraction rules composed only of simple contextual patterns. Extraction is triggered by specific token sequences preceding and following the start or end of a target field.

## 2  Information Extraction

Past work has focused mainly on the construction and learning of the extraction rule. The rule must be applicable to several fields while making few errors and extracting the maximum of relevant information in the document or web page. However, documents such as web pages are semi structured and present several anomalies. Any particular field in a web page may present a varying structure as well as a variant context. Hence, it is difficult to construct a perfect rule that satisfies all conditions. The construction of a complex rule or the adoption of a complex learning algorithm does not resolve the difficulty. The approach we propose uses a simple extraction rule that targets a single field. This has enabled us to design a reasonable simple learning algorithm. The extraction rule exploits the expressive power of fuzzy algebra to accommodate variant anomalies that may be present in the field, which make our approach extremely flexible.

An extraction mask is a procedure that extracts useful information (in response to a user request) contained in a given document. The extracted information is then produced in a structured format defined by the mask. The mask shows useful information, i.e. useless information is hidden from the user.

Several approaches have been proposed to help construct extraction masks. Some are completely manual, while others are automatic or semi automatic (N.Ashish, C. Knob lock., 1997).

Manual approaches to mask construction describe the web structure with grammars. This approach requires expert intervention to design the appropriate grammar as well as to maintain the mask when the structure of the information source changes.

For semi automatic approaches the user instructs the system, via an interface, which information fields to extract. The system then constructs the adequate mask. These approaches do not require the intervention of an expert. However, any change in the structure of the information source implies user intervention. Our approach is semi automatic and generates masks by induction.

Automatic approaches employ learning techniques based on the use of heuristics, case based reasoning, etc.

Inductive learning algorithms proceeds either *bottom-up* (generalisation) or top-down (specialisation) (D. Freitag, N. Kushmerick, 2000). A bottom-up approach starts by selecting one or several examples and constructing a hypothesis to cover them. Next, it tries to generalise the hypothesis to cover the rest of the examples. On the other hand, a top-down approach starts with a general hypothesis and then tries to refine it to cover all positive examples and none of the negative examples.

The information extraction is a complex process. It consists both of a learning task and an extraction task. Most IE systems have the architecture given in Figure 1.

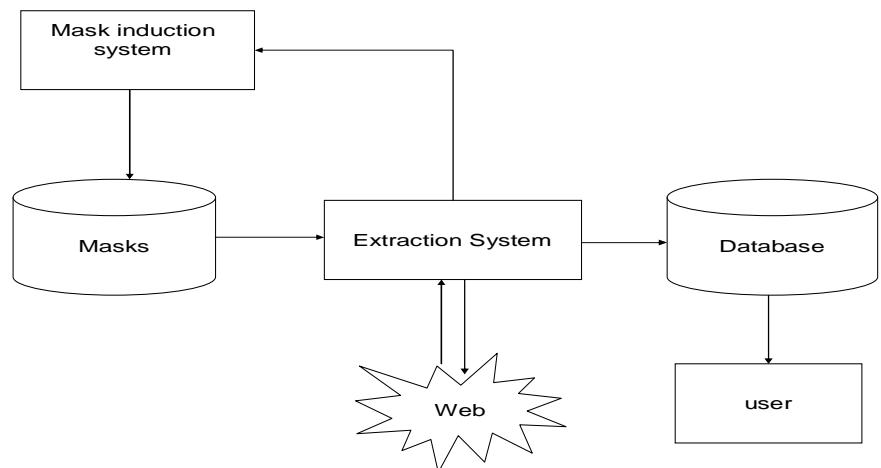

**Fig. 1.** Architecture of an IE system.

In general, a web page is composed of a sequence of tokens. A token may take the form of a simple character, an html tag, a string of digits, etc.

A Web page consists of three main zones: a **global zone**, a **record zone**, and an **attribute zone**. The **global zone** contains the various tuples of the page. The **record zone** consists of the tuple to extract. The **attribute** zone is the text fragment sought and is encapsulated in the tuple (see Figure 2).

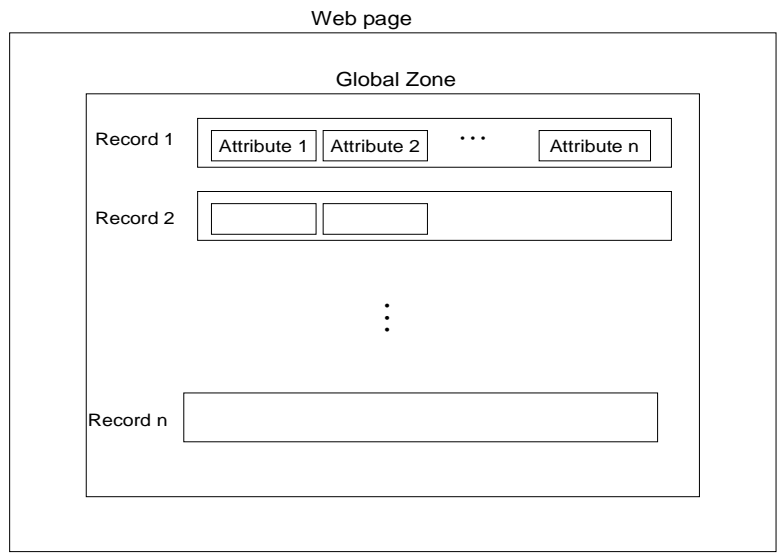

**Fig. 2.** Architecture of a Web page.

A zone is marked by a Begin Separator of the zone and an End Separator of the zone. A separator is composed of two token sequences, DetectorL and DetectorR. Each sequence is called Detector. Figure 3 illustrates the structure of a zone.

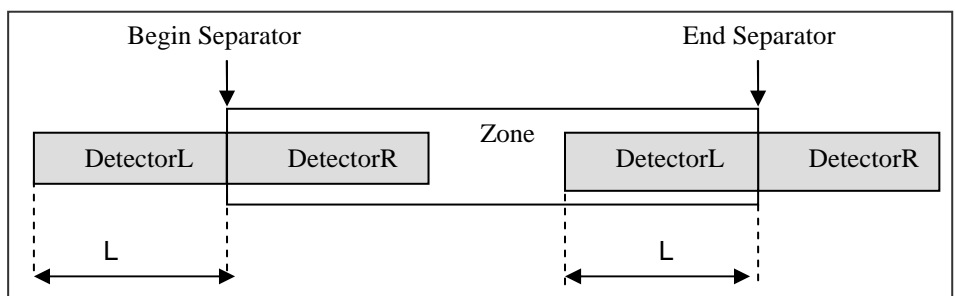

**Fig. 3.** Structure of a zone. L is the average length of a tuple.

## 3 Related works

Four of the most popular wrappers for information extraction are WIEN (Wrapper Induction Environment) (N.Kushmerick., 1997; and N.Kushmerick, D.Weld, R.Doorenbos. 1997), BWI (Boosted Wrapper Induction) (D. Freitag, N. Kushmerick, 2000), WHISK (S. Soderland. 1996; S. Soderland. 1999), and SoftMealy (C. Hsu and M. Dung., 1998; C-N.Hsu, 1998).

WIEN (D. Freitag, N. Kushmerick, 2000; N.Kushmerick., 1997; and N.Kushmerick, D.Weld, R.Doorenbos. 1997) is a wrapper which automatically constructs masks based on a user supplied web page examples. WIEN is capable of information extraction from web pages having an array format.

Several classes of masks have been designed enabling the extraction of the various tuples that are present in a web page. A mask is composed of couples of strings delimiting the attributes of a tuple. Hence, each attribute containing the desired text fragment is delimited by a left delimiter and a right delimiter. The mask induction algorithm repetitively generates an extraction mask and tests it on the user supplied examples until it finds a mask that covers all the examples.

The extraction process consists of locating for each attribute its left and right delimiters and extracting the information between the two delimiters.

BWI (D. Freitag, N. Kushmerick, 2000; D. Freitag., 1998; D. Freitag., 2000) is a mono-attribute trainable information extraction system. The extraction algorithm learns separately two sets of boundary detectors: a set F to detect the start boundaries of the desired attribute and a set A to detect the end boundaries of the attribute. The learning attribute associates with each learned detector a confidence value, which is a function of the number of examples correctly covered and the number of miscovered examples. The confidence values are used to compute a weight for each example. The weights allow the definition of the learning rate of the supplied examples. Then, examples with low weights are considered not well learned and are given preferential treatment over examples with large weights, which are considered well learned. Extraction consists of seeking begin and end separators with an error below a given minimum error.

WHISK (S. Soderland. 1996; S. Soderland. 1999) is designed to handle text styles ranging from highly structured to free text, including text that is neither rigidly formatted nor composed of grammatical sentences. Such semi-structured text has largely been beyond the scope of previous systems. When used in conjunction with a syntactic analyzer and semantic tagging, WHISK can also handle extraction from free text such as news stories.

SoftMealy (C. Hsu and M. Dung., 1998; C-N.Hsu, 1998) is a multi-attribute extraction system based on a new formalism of mask representation. This representation is based on a Finite-State Transducer (FST) and contextual rules, which allow a wrapper to wrap semi-structured Web pages with missing attributes, multiple attribute values, variant attribute permutations, exceptions, and typos. The nodes (states) of the FST model the zones of the Web page and the transitions the possible zone separators. The FST in SoftMealy takes a sequence of the separators rather than the raw HTML string as input. Each distinct attribute permutation in the Web page can be encoded as a successful path and the state transitions are determined by matching contextual rules that describes the context delimiting two adjacent attributes.

## 4 A new fuzzy wrapper for Information Extraction

The solution we propose is based on a new formalism for rule extraction and uses the expressive power of fuzzy logic during the process of extraction.

Similar to the SoftMealy wrapper, before we start the extraction of the tuples, we segment an input HTML string into tokens. A token is denoted as t(v) where t is a token class and v is a string. For example, to the html tags "<I>" and "<B>" correspond the tokens Html (<I>) and Html (<B>), and to the numeric string "123" correspond the token Num (123). Below are the token classes and their examples.

- All uppercase string : $"FSM" \rightarrow CAlph(FSM)$
- An uppercase letter, followed by string with at least one lower case latter : $"Professor" \rightarrow C1Alph(Professor)$
- A lower case letter followed by zero or more characters: $"and" \rightarrow 0Alph(and)$
- Numeric string : $"123" \rightarrow Num(123)$
- An opened HTML tag : $"<I>" \rightarrow Html(<I>)$
- A closed HTML tag : $"</I>" \rightarrow /Html(</I>)$
- Punctuation symbol : $"," \rightarrow Punc(,)$
- An opened HTML tag representing control characters, $a\ new\ line \rightarrow NL(1)$, $two\ blanc\ spaces \rightarrow Spc(2)$
- A closed HTML tag representing control characters
- An opened HTML tag representing element of a list
- A closed HTML tag representing element of a list
- A generic string *Any* representing any other class

Our wrapper is capable to learn how to extract structure information from web resources on the basis of user supplied examples. It is based on inductive learning techniques as well as fuzzy logic rules. This approach is composed of three modules: a page labelling module, a learning module, and an extraction module.

The first module is the page labelling one. The main task of this module is the specification of the Web page architecture. It indicates the beginning and end of each zone in the page. The module interacts with the user for the specification of the zone boundaries. This module accepts as input a Web page and produces as output a series of labels.

The learning module takes as input a web page and its labels and constructs extraction rules for each zone. The learning of a zone consists of the determination of the extraction rule that will recognize the two separators at the beginning and end of each zone.

To recognize the separator at the beginning or end of a zone, we must identify the pertinent tokens among the token sequence to the left of the separator (DetectorL) and those to its right (DetectorR).

The learning step consists of determining in the token sequence the positions of the pertinent tokens and their occurrences over a distance MoyL, which is the tuple average length. Then, for each detector we determine a frequency matrix $F$ where, $f(i,j)$ represents the number of occurrences of token j at distance i.

Next, we estimate the cost of each detector. The cost metric is used to estimate the error made by a detector that is learned as opposed to a detector that is extracted. This metric should be a function of the token positions and their occurrences.

Let P and O be, respectively, the set of the positions and the occurrences of learned tokens. To estimate the cost of a token, we define the following two functions:

$f_P^j : \mathbb{N} \to [0,1]$: is a function characterizing the degree of truth of the position of the token j with respect to its separator (distance), defined as follows (see Figure 6):

$$\begin{cases} f_P^j(i) = 1 & \text{if} \quad f(i,j) > 0 \\ 0 \leq f_P^j(i) < 1 & \text{if} \quad f(i,j) = 0 \end{cases} \quad . \qquad (1)$$

$f_O^j : \mathbb{N} \to [0,1]$: is a function characterizing the degree of truth of the occurrence of the token j. It is defined as follows:

$$\begin{cases} f_O^j(i) = \dfrac{f(i,j)}{\text{number of learned instances}} & \text{if} \quad f(i,j) > 0 \\ f_O^j(i) = \dfrac{f(i',j)}{\text{number of learned instances}} & \text{ifi} \quad f(i,j) = 0 \end{cases} \quad . \qquad (2)$$

With $i'$ is the nearest position of $i$ such as $f(i,j) > 0$.

The cost of a token and the cost of a detector are defined as follows:

$$\text{Cost}_{\text{token}}(\text{distance}) = f_P^{\text{token}}(\text{distance}) \cdot f_O^{\text{token}}(\text{distance}) . \qquad (3)$$

$$\text{Cost}_{\text{detector}} = \sum_{i=1}^{\text{moyL}} \text{Cost}_{\text{token}}(i) \qquad (4)$$

Then, while parsing the web pages, we associate with each zone detector the minimum cost of the encountered detector $C_{min}$, the maximum cost $C_{max}$, and the average cost,

$$C_{moy} = \frac{C_{min} + C_{max}}{2}. \quad (5)$$

These cost metrics will serve during the following extraction stage.

The third module is the extraction module consists of extracting the different tuples contained in a Web page.

To extract the different tuples a Web page, we proceed in three steps. First, we extract the global zone of the page. Next, the various records contained in the global zone are extracted. Finally, for each record, we extract the different attributes it contains to construct the corresponding tuple. This way, all tuples of the page are extracted.

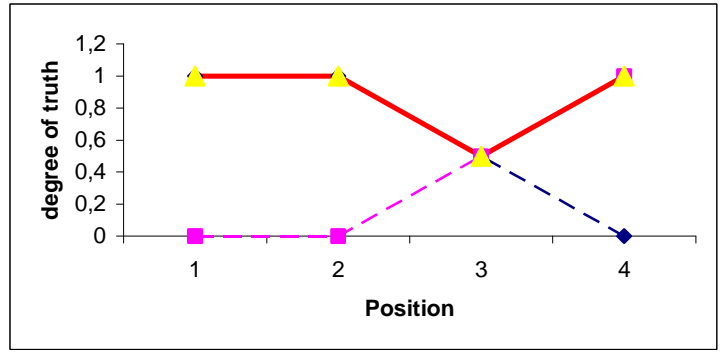

**Fig 4.** Function specifying the degree of truth of the Position of token C1Alph.

The extraction of a zone is done via the determination of the two separators of the beginning and end of the zone. The determination of a separator is achieved by means of its two detectors. Indeed, the separator we seek is the one whose two detectors commit minimal cost errors (less than a threshold) in comparison to the costs of the detectors learned during the learning stage.

$$. \, Cost_{Detector} = Cost_{Detector} - C_{moy} \quad (6)$$

$$Error_{Separator} = Error_{DetectorL} + Error_{DetectorR} \quad (7)$$

To determine the error that a separator commits from the errors committed by its detectors DetectorL and DetectorR, we apply the following fuzzy logic rules.

- if ErrorLeft is positiveSmall or ( ErrorRight is positiveSmall) then ErrorTot is positiveSmall.
- if ErrorLeft is positive or ( ErrorRight is positive) then ErrorTot is positive.
- if ErrorRight is zero and ( ErrortLeft is zero ) then ErrorTot is zero.
- if ErrorLeft is negativeSmall or ( ErrorRight is negativeSmall) then ErrorTot is negativeSmall.
- if ErrorLeft is negative or (ErrorRight is negative) then ErrorTot is negative.

## 5 Exemple

Let's consider web pages listing country names and codes (Figure 5). Each tuple consists of a country name and its corresponding telephone code. To label a Web page, we use an interface that allows a user to specify for each zone the starting and ending characters of the zone.

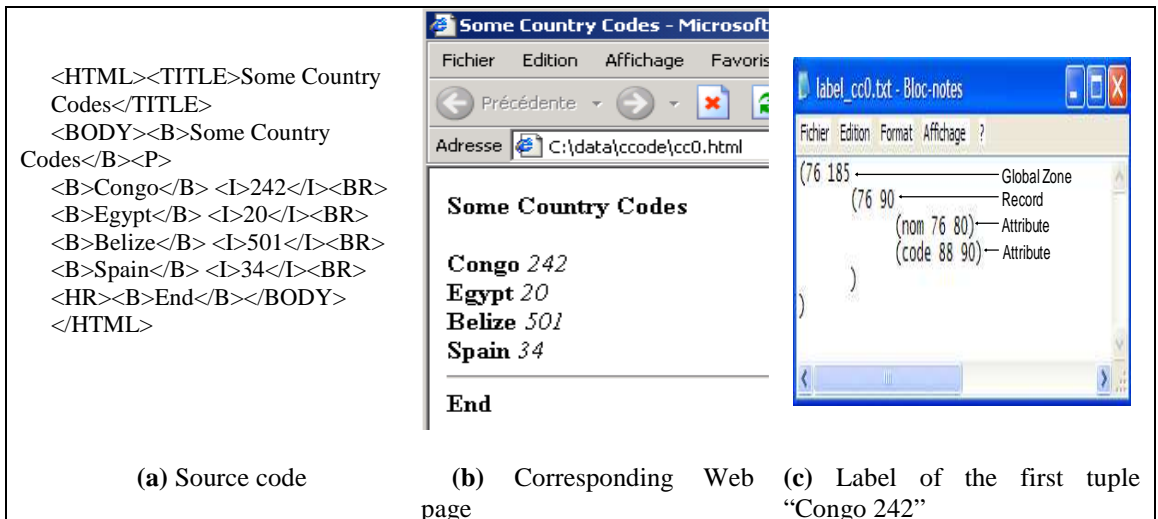

(a) Source code  (b) Corresponding Web page  (c) Label of the first tuple "Congo 242"

**Fig. 5.** Labelling of a Web page.

The learning module accepts as input a sequence of couples (Web Page, Label). In this example, we used three pages and, therefore, the learning module is given three couples. Then, the information extraction system builds for each detector a frequency matrix. For the Web page example of Figure 5, the frequency matrix obtained after the learning of DetectorL of the separator of the start of the global zone is given in Table 1.

For example, FrequencyL(4,1) = 3 means that the token C1Alph has been observed in three pages at position 4 in DetectorL. Then we compute for each detector the corresponding costs, $C_{min}$, $C_{max}$ et $C_{moy}$. The cost of a detector is equal to the sum of the costs of its tokens.

|          |    | Tokens |       |     |       |      |      |     |      |     |       |      |     |
|----------|----|--------|-------|-----|-------|------|------|-----|------|-----|-------|------|-----|
|          |    | C1Alph | CAlph | Num | 0Alph | Punc | /Spc | Spc | /Lst | Lst | /Html | Html | Any |
|          |    | 01     | 02    | 03  | 04    | 05   | 06   | 07  | 08   | 09  | 10    | 11   | 12  |
|          | 00 | 0      | 0     | 0   | 0     | 0    | 0    | 0   | 0    | 0   | 0     | 0    | 0   |
| Distance | 01 | 2      | 0     | 0   | 0     | 0    | 0    | 0   | 0    | 0   | 0     | 1    | 0   |
|          | 02 | 2      | 0     | 0   | 0     | 1    | 0    | 0   | 0    | 0   | 0     | 0    | 0   |
|          | 03 | 0      | 0     | 0   | 0     | 0    | 0    | 0   | 0    | 2   | 0     | 0    | 1   |
|          | 04 | 3      | 0     | 0   | 0     | 0    | 0    | 0   | 0    | 0   | 0     | 0    | 0   |

**Table 1:** FrequencyL, the frequency matrix of DetectorL of the separator of the start of the global zone.

For example, suppose that DetectorL has the following structure:

|             | C1Alph | C1Alph | Lst | HTML |
|-------------|--------|--------|-----|------|
| **Distance :** | 4      | 3      | 2   | 1    |

To compute the cost of the token C1Alph, we look at the first column of matrix FrequencyL. The degree of truth function corresponding to the position of a token is given in Figure 4.

Then the costs corresponding to the token C1Alph are,

$$C_{C1Alph}(4) = 1*3/3 = 1$$
$$C_{C1Alph}(3) = 0.5*3/3 = 0.5$$

and the cost of the detector is,

$$C_{détecteur} = C_{C1Alph}(4) + C_{C1Alph}(3) + C_{Lst}(2) + C_{HTML}$$
$$= 1 \ * \ 3/3 + 0.5 \ * 3/3 + 0.66 \ * \ 2/3 + 1 \ * \ 1/3$$
$$= 2.26$$

## 5 Experimental results

We compared performances of our approach to those of SoftMealy. We considered five collections of Web pages that present different types of anomalies and attempted to extract from each page the different tuples it contains.

The results are summarised in Table 2 below. We compare our approach with SoftMealy, since only the code of SoftMealy was available to us. The comparison is performed with respect to the *Recall Coefficient* and *Precision* performance metrics, which are defined as follows:

$$Recall = \frac{number\ of\ extracted\ tuples}{total\ number\ of\ tuples\ in\ the\ Web\ page}. \quad (8)$$

$$Precision = \frac{number\ of\ extracted\ pertinent\ tuples}{total\ number\ of\ tuples\ in\ the\ Web\ page} \quad (9)$$

|  | Set of Web pages<br>Results | Set_1<br>(5 pages) | Set_2<br>(11 pages) | Set_3<br>(17 pages) | Set_4<br>(23 pages) | Set_5<br>(33 pages) |
|---|---|---|---|---|---|---|
|  | Total number of tuples | 20 | 35 | 55 | 73 | 108 |
| **SoftMealy** | Number of extracted tuples | 11 | 30 | 12 | 14 | 20 |
| | Number of pertinent tuples extracted | 11 | 20 | 7 | 10 | 19 |
| **our approach** | Number of extracted tuples | 12 | 34 | 20 | 30 | 33 |
| | Number of pertinent tuples extracted | 10 | 23 | 11 | 18 | 26 |

**Table 2.** Number of extracted tuples pertinent or otherwise obtained with Soft Mealy and our approach for each of the test pages.

The diagram of the figure 6 summarizes results shown in the table 2. We notice that the number of tuples retrieved by our approach is superior to those retrieved by SoftMealy. The efficiency of our approach increases advantage when the number of pages increases from a set to another.

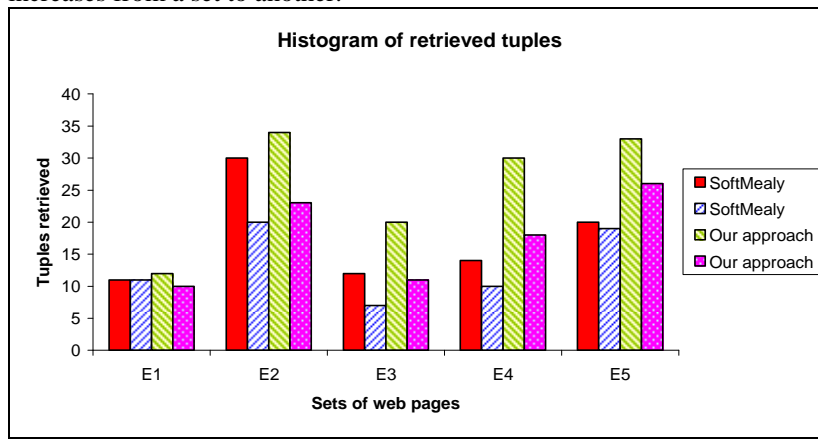

**Fig. 6.** Histogram of obtained results

Figures 7 and 8 plot the *Recall Coefficient* and *Precision* performance measures, obtained by the SoftMealy wrapper and by our approach, for each of the Sets.

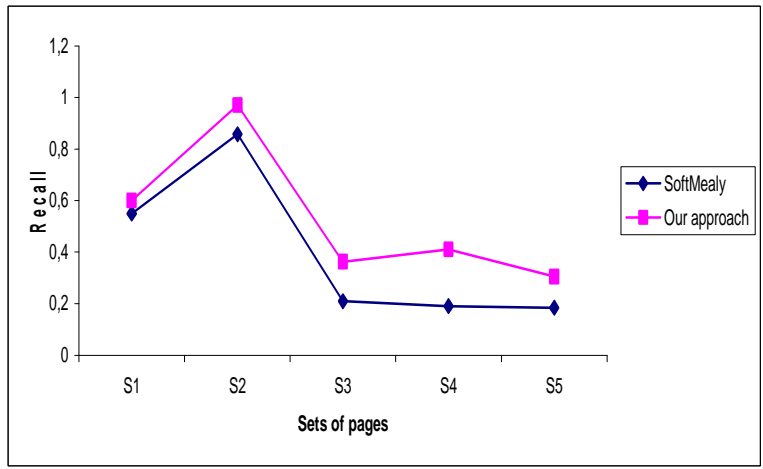

**Fig. 7.** Comparison between SoftMealy and our approach with respect to the Recall Coefficient metric.

We notice that, with regard to the recall coefficient, the two curves corresponding to the two approaches follow the same pace. This phenomenon can be explained by the fact that processes of training are the same in the two approaches. However, figure 7 and 8 clearly illustrate that the recall coefficient of our approach is always superior to the recall coefficient reached by SoftMealy and the gap between the two approaches is more and more important when the number of test pages increases.

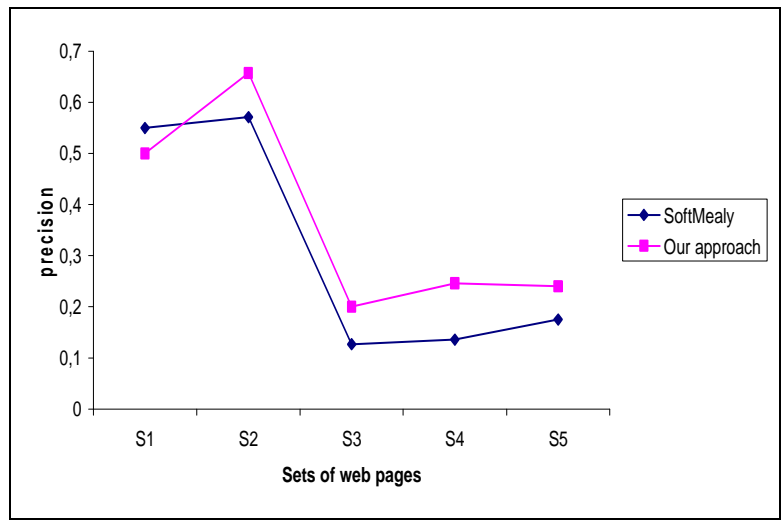

**Fig. 8.** Comparison between SoftMealy and our approach with respect to Precision metric.

We notice, on the other hand, that the precision given by our approach is superior to the one of SoftMealy although they nearly have the same pace. The precision given by SoftMealy is slightly better that the one of our approach, when we consider small sets of pages web.

## 6   Conclusion

In this paper, we presented a new approach to information extraction from multi-attribute semi-structured Web pages. Our approach is flexible in the sense that it tolerates the existence of variant anomalies in the pages such as, missing attributes, permutation of attributes, etc. The flexibility is achieved by following a fuzzy inductive learning approach. The user can intervene at any moment to improve the rules learned by adding new training examples. Both the information extraction and learning algorithms are independent of the lexical analyser. Experimental results obtained on several test Web pages show a superior performance of our approach compared to that of SoftMealy with respect to Recall Coefficient and Precision metrics.

The basis on which we worked is characterized by different exceptions. SoftMealy, for example, is very dependent of the training whole. Indeed, transitions of the automaton are very dependent on what is observed during the training. The function of generalization used in SoftMealy permits no mistake of positions for tokens. It doesn't hold amount of the token occurrence in the whole of training. So, a transition remains rigid enough even though it is constituted of several under rules.

To determine the beginning and the end of a zone, our approach uses the committed mistake. It permits to assign a mistake to the observed processes during the training. It permits to detect, as SoftMealy, the different permutations of attributes. The resemblance of our approach with SoftMealy is to the level of the exploitation of the same architecture of a Web page during the phase of extraction.